\documentclass{iau}
\usepackage{graphicx}

\title[Kinematics of M dwarfs in CARMENES ] 
{Kinematics of M dwarfs in the CARMENES input catalogue: \\   membership in young moving groups}

\author[D.~Montes et al.]   
{
D.~Montes$^{1}$, 
J. A. Caballero$^{2}$, 
I. Gallardo$^{1}$,
M. Cort\'es-Contreras$^{1}$, 
 \and F. J. Alonso-Floriano$^{1}$
 }

\affiliation{$^1$Dpto. Astrof\'{\i}sica, Facultad de CC. F\'{\i}sicas, Universidad Complutense de Madrid, E-28040 Madrid, Spain\\email: {\tt dmontes@ucm.es}\\
$^2$Centro de Astrobiolog\'{\i}a (CSIC-INTA), Campus ESAC, PO Box 78, E-28691 Villanueva de la Ca\~{n}ada, Madrid, Spain
}

\pubyear{2015}
\volume{314}  
\pagerange{119--126}
\jname{Young Stars \& Planets Near the Sun}
\editors{J. H. Kastner, B. Stelzer, \& S. A. Metchev, eds.}
\begin{document}

\maketitle

\begin{abstract}
We present a detailed study of the kinematics of M dwarfs in the 
CARMENES (Calar Alto high-Resolution search for M dwarfs with Exoearths with Near-infrared and optical \'Echelle Spectrographs) input catalog. 
We have selected all M dwarfs with known parallactic distance or a good photometric distance estimation, 
precise proper motion in the literature or as determined by us, and radial velocity measurements.
Using these parameters, we computed the M dwarfsÕ galactic space motions ($U$, $V$, $W$).
For the stars with $U$ and $V$ velocity components inside or near the boundaries that determine the young disk population, we have analyzed the possible membership in the classical moving groups and nearby loose associations with ages between 10 and 600 Ma. For the candidate members, we have compiled information available in the literature in order to constrain their membership by applying other age-dating methods.
\end{abstract}

\vspace*{-0.2 cm}
\firstsection 
\section{Introduction}

We are compiling the most comprehensive database of M dwarfs ever built, CARMENCITA, the CARMENES Cool dwarf Information and daTa Archive, 
which will be the CARMENES 'input catalog' (\cite{Quirren14}).  
In addition to the science preparation with low- and high-resolution spectrographs and lucky imagers 
(see \cite{AF15}; \cite{CC14}; \cite{Passegger15}), 
we have compiled a large amount of public data on over 2100 M dwarfs, and we are analyzing them.
Here we describe the preliminary results about the kinematics derived from these data.

\begin{figure}[t]
 \vspace*{-0.2 cm}
\begin{center}
\centerline{
\includegraphics[width=5.4in]{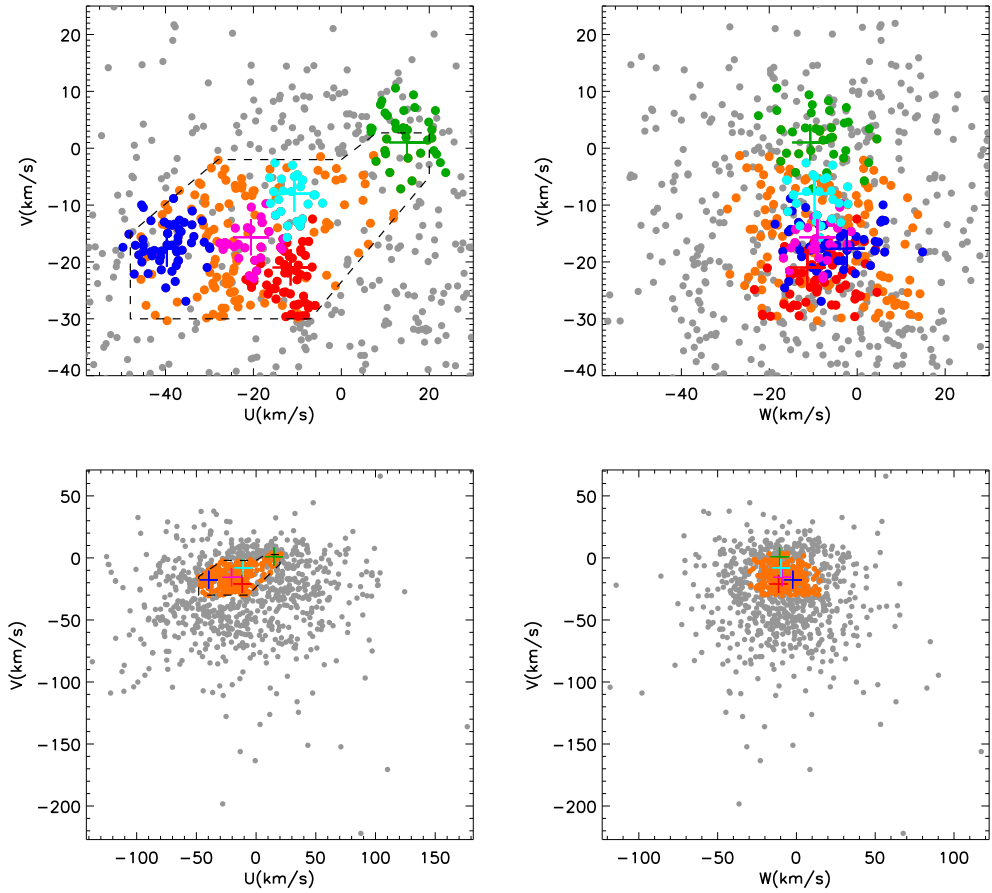}
}
 \vspace*{-0.2 cm}
 \caption{
\textit{Upper panel}: a zoom in $UV$ and $WV$ diagrams around the boundaries (dashed line) that determine the young disk population 
as defined by \cite{Eggen89} 
and the position of the classical moving groups (Local Association, Hyades Supercluster, Ursa Major, Castor, IC 2391 and Hercules-Lyra). The different colors indicate the possible candidates to each moving group.
\textit{Lower panel}: The full range of motions on $UV$ and $WV$ diagrams.}
   \label{fig:uvw}
\end{center}
\end{figure}

\vspace*{-0.4 cm}
\section{Membership in young moving groups}

From CARMENCITA we have now 1462 M dwarfs with all  parameters (distance, radial velocity and proper motion) 
needed to determine the Galactic velocity components ($U$, $V$, $W$). 
When parallactic distance is not available we have derived a photometric distance. 
For a large group of stars (in particular 214 stars in \cite{LPSM}, LSPM) 
with 
imprecise proper motions or missing uncertainties 
in the literature we have determined the proper motions
using data from 2MASS, CMC14, CMC15, GSC2.3, USNO-A2, SDSS-DR9, ALLWISE and SuperCOSMOS.
Figure~\ref{fig:uvw} shows the ($U$, $V$, $W$) diagrams for the 1462 M dwarfs in CARMENCITA with accumulated data 
%
and the possible candidate members of moving groups (\cite{2001MNRAS.328...45M}; \cite[Montes 2010, 2015]{mon10, mon15}; \cite{Klutsch14}). 

\begin{acknowledgements}
This work was supported by the Univ. Complutense de Madrid (UCM) and the Ministry of Economy and Competitiveness (MINECO) under grant AYA2011-30147-C03-02.
\end{acknowledgements}


%


\end{document}